\documentclass[showpacs,prl,twocolumn,aps]{revtex4}

\usepackage{hyperref}
\usepackage{amsmath}
\usepackage{amssymb}
\usepackage{latexsym}
\usepackage{amsfonts}
\usepackage{epsfig}
\usepackage{psfrag}
\usepackage{graphicx}

\usepackage{color}

\newcommand{\bea}{\begin{eqnarray}}
\newcommand{\ea}{\end{eqnarray}}
\newcommand{\eea}{\end{eqnarray}}

\begin{document}

\title{The Stokes Phenomenon  and Schwinger Vacuum Pair Production in Time-Dependent Laser Pulses}

\author{Cesim~K.~Dumlu and Gerald~V.~Dunne}

\affiliation{Department of Physics, University of Connecticut, 
Storrs CT 06269-3046, USA}

\begin{abstract}
Particle production due to external fields (electric, chromo-electric or gravitational) requires evolving an initial state through an interaction with a time-dependent background, with the rate being computed from a Bogoliubov transformation between the in and out vacua. When the background fields have  temporal profiles with sub-structure, a semiclassical analysis of this problem confronts the full subtlety of the Stokes phenomenon: WKB solutions are only local, while the production rate requires global information. Incorporating the Stokes phenomenon, we give a simple quantitative explanation of the recently computed  [Phys. Rev.\ Lett.\ {\bf 102}, 150404 (2009)] oscillatory momentum spectrum of $e^+\,e^-$ pairs produced from vacuum subjected to a time-dependent electric field with sub-cycle laser pulse  structure. This approach also explains  naturally why  for spinor and scalar QED these oscillations are out of phase. 
\end{abstract}



\maketitle

The Schwinger effect, the non-perturbative production of electron-positron pairs from vacuum in an external electric field, is a highly non-trivial prediction of QED \cite{eh,schwinger,dunne-eli}, but the physical scales are such that it is so weak that it has not yet been directly observed. Recent experimental advances  \cite{tajima} have raised hopes that lasers may achieve fields  just one or two orders of magnitude below the critical field strength of $E_{\rm cr}\sim 10^{16}$ V/cm, either in optical high-intensity laser facilities such as HiPER (Rutherford Lab.) and the Extreme Light Infrastructure (ELI), or in X-ray free electron laser facilities. Theoretically, recent analyses suggest that the non-perturbative Schwinger effect may be observable at these lower field strengths, by careful shaping and combining of laser pulses leading to a "dynamically assisted Schwinger mechanism'' enhancement \cite{Schutzhold:2008pz,Bell:2008zzb,DiPiazza:2009py,Monin:2009aj,Heinzl:2010vg,Bulanov:2010ei}. The most important message is that the detailed shape of the laser pulse is significant, which motivates the extension presented here of the standard WKB approach to include more realistic laser field profiles. 

Observation of the Schwinger effect in the non-perturbative domain has the potential to yield valuable insight into analogous gravitational effects \cite{ralf-eli,leonhardt}, such as Unruh and Hawking radiation, where direct experiments are not feasible, and where issues such as back-reaction and out-of-equilibrium physics are poorly understood. The basic physics is also relevant for atomic, molecular, astro- and plasma  physics with ultra-high intensity lasers, where non-perturbative effects are crucial \cite{mourou,astro}, for heavy ion collisions \cite{dima},  and for the Landau-Zener effect.

We model the electric field  in the focal region of two counter-propagating laser pulses by a spatially homogeneous electric field $\vec{E}(t)=(0, 0, E(t))$, with vector potential $\vec{A}(t)=(0, 0, A(t))$, such that $E(t)=-\dot{A}(t)$. Even in this approximation where we neglect spatial focussing, the laser field may involve many physical scales, leading to interesting new phenomena \cite{Hebenstreit:2009km}. For example, consider an electric field:
\begin{eqnarray}
  \label{eqn:elfield}
   E(t)=E_0\cos(\omega t+\phi)\exp\left(-t^2/(2\tau^2)\right)\ .
\end{eqnarray}
Here $\omega$ is the laser frequency, $\tau$ defines the pulse length, and $\phi$ is the ``carrier phase'' offset. The first surprising result in \cite{Hebenstreit:2009km} was that the longitudinal  momentum spectrum of the produced electron-positron pairs is extremely sensitive to the value of $\omega \tau$, even when $\phi=0$. For $\omega \tau\gtrsim 4$, the momentum spectrum exhibits oscillations, and these become dramatically enhanced as $\phi$ increases, to the point where at $\phi=\pi/2$ the spectrum develops minima with zero produced pairs (see Fig. 4 in  \cite{Hebenstreit:2009km}). The second surprising result in \cite{Hebenstreit:2009km} was that  the oscillatory minima and maxima are interchanged between spinor and scalar QED. By  contrast, when computing the {\it total} pair production rate (obtained by an integral over the momenta),  one conventionally approximates the case of real spinor QED by scalar QED, with an overall multiplicative spin factor of 2. A direct application of the usual semiclassical "imaginary time method" (ITM) \cite{brezin,popov,popov-ionization,kimpage,Kleinert:2008sj} to this problem does not account for these oscillations, let alone for the difference of phase between spinor and scalar QED. Here we show that the Stokes phenomenon gives a {\it quantitative} semiclassical explanation of  both these effects.
 
The essential physical interpretation of these oscillations is a resonance effect in the corresponding quantum mechanical scattering problem \cite{Hebenstreit:2009km}. This same physical explanation has also been noted for the photoelectron spectrum in atomic  ionization \cite{popov-ionization}, where such oscillations have been observed \cite{doubleslit,paulus}. We turn this physical picture into a {\it quantitative} method. This should also be relevant for the matterless double-slit experiment \cite{Hebenstreit:2009km,king}.
Recall that with a time-dependent electric field, the pair production process can be reduced to a one-dimensional  over-the-barrier "quantum mechanical" scattering problem \cite{brezin,popov,dunne-eli}, with effective "Schr\"odinger equation" [in $t$ rather than $x$]
\begin{eqnarray}
\ddot \phi+Q^2(t)\,\phi=0\quad , \quad Q^2(t)\equiv m^2+p_\perp^2+\left(p-A(t)\right)^2
\end{eqnarray}
coming from the Klein-Gordon  equation for the particle  in the presence of the laser field.  (We first discuss scalar QED and later come to spinor QED, where the relevant equation is the Dirac equation.) It is an over-the-barrier scattering problem since the "potential", $-(p-A(t))^2$, is negative, while the "energy" $(m^2+p_\perp^2)$ is positive. Thus the reflection coefficient, from which we deduce the probability of pair production, is exponentially small in the semiclassical regime where $E_0 \ll m^2$. This scattering problem can be solved using WKB methods, or numerically using the quantum kinetic approach \cite{kluger,Hebenstreit:2009km} or direct integration of the scattering problem \cite{popov}. The equivalence between these approaches is explained in \cite{Dumlu:2009rr}. 
 
 The Bogoliubov transformation approach \cite{brezin,popov,popov-ionization,kimpage,Kleinert:2008sj,kluger} is based on the field decomposition
 \begin{eqnarray}
 \phi&=&\frac{\alpha}{\sqrt{2 Q}}e^{-i\int^t Q}+\frac{\beta}{\sqrt{2 Q}}e^{i\int^t Q}
 \nonumber\\
 \dot\phi&=&-iQ\left(\frac{\alpha}{\sqrt{2 Q}}e^{-i\int^t Q}-\frac{\beta}{\sqrt{2 Q}}e^{i\int^t Q}\right)
 \label{decomp}
 \end{eqnarray}
which enforces equations relating the coefficient functions $\alpha$ and $\beta$. Unitarity requires $|\alpha|^2-|\beta|^2=1$, and the particle number momentum spectrum  is $N(\vec{p})=|\beta_{\vec{p}}(t=\infty)|^2$,  related to the reflection coefficient  $|R|^2=|\beta|^2/(1+|\beta|^2)$. The Stokes phenomenon is relevant because in calculating $|R|^2$ we compare $\beta(t=\infty)$ to $\beta(t=-\infty)$. However, the leading WKB solutions, $e^{\pm i \int Q}/\sqrt{2 Q}$, on which (\ref{decomp}) is based, are multi-valued functions, only defined {\it locally}. Evolving a semiclassical approximation from $t=-\infty$ to $t=+\infty$, we cross Stokes and anti-Stokes lines,  lines along which $e^{\pm i\int^t Q}$ are exponential or oscillatory, respectively. On crossing such lines, we must take care to keep track properly of the dominant and sub-dominant solutions. This is the Stokes phenomenon \cite{heading,Berry:1972na,bender,white}. 

Suppose  the zeros of $Q(t)$, the ``turning points'' (t.p.'s), are first-order. Since this is an over-the-barrier scattering problem, the t.p.'s lie off the real axis, in the complex plane, and for real laser pulses they occur in complex conjugate pairs; furthermore, those closest to the real axis tend to dominate in the semiclassical regime. For a simple single-pulse field  
like $E(t)=E\, {\rm sech}^2(\omega t)$, with $A(t)=-E/\omega \tanh(\omega t)$, or $E(t)=E\, e^{-(\omega t)^2}$, with $A(t)=-\sqrt{\pi}E/(2\omega) \, {\rm Erf}(\omega t)$, a single pair of complex conjugate t.p.'s dominates, and the phase integral method leads to the familiar formula \cite{landau,pokrovskii,heading,popov-ionization,white}
\begin{eqnarray}
N(\vec{p}) \approx e^{-2 K}\quad , \quad K=\left| \int_{tp} Q dt\right|
\label{onepair}
\end{eqnarray}
where the integral is along the line joining the two complex conjugate t.p.'s. In the ITM, one  expands in momenta to obtain the general Gaussian expression \cite{popov}
 \begin{eqnarray}
N(\vec{p}) &\approx& \exp\left[-S_{\rm cl} -c_1\, p_\perp^2 -c_2 \, p^2\right] 
 \label{gaussian}
 \end{eqnarray}
 where $S_{\rm cl}$ is the classical action evaluated on the contour,  $c_1=\frac{\partial}{\partial m^2} S_{\rm cl}$, and $c_2=-2 m^2 \frac{\partial^2}{\partial (m^2)^2} S_{\rm cl}$.  This ITM  result (\ref{gaussian}) \cite{popov,popov-ionization} gives a compact expression that is the basis for most studies of vacuum pair production, and when integrated over momentum to give the total rate it gives excellent agreement with numerical (or exact) results \cite{popov,kimpage,dunnehall,dunne-eli,florian-1,Kleinert:2008sj}. In \cite{popov-pulse}, (\ref{gaussian}) was applied to the envelope field (\ref{eqn:elfield})  with $\phi=0$.  However, the expression (\ref{gaussian}) clearly cannot exhibit any of the numerically observed  oscillations, as noted in \cite{Hebenstreit:2009km}. This deficiency is not cured by  higher order terms in the momentum expansion in (\ref{gaussian}), nor is it cured by including higher-order WKB terms. The problem is that (\ref{onepair}) and (\ref{gaussian}) are based on the assumption of just one pair of  turning points, on the imaginary axis. However, for complicated fields, with sub-cycle structure,  the essential shape of the "scattering potential", $-(p-A(t))^2$,  changes dramatically as $p$ varies, as illustrated in Fig 1. This can lead to scattering resonances, encoded semiclassically in {\it multiple} pairs of complex conjugate t.p.'s, whose locations are correspondingly sensitive to variation of $p$. We show below that the  oscillatory momentum behavior can be identified with interference effects between such pairs of turning points.
\begin{figure}[htb]
\includegraphics[scale=0.7]{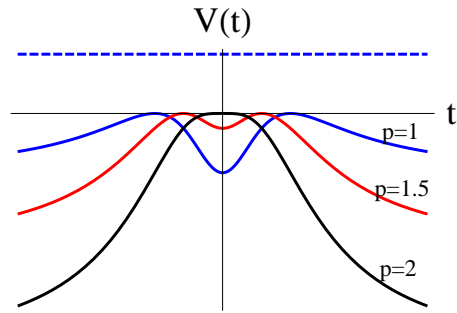}
\caption{The effective scattering potential, $V(t)=-(p-A(t))^2$, for  scalar QED,  with the field in (\ref{evenfield}), for three different values of the longitudinal momentum, in units of m. Note that the form of the potential is highly sensitive to the value of $p$. The dashed line denotes the mass level $m^2$.}
\label{potential}
\end{figure}

\begin{figure}[htb]
\includegraphics[scale=0.55]{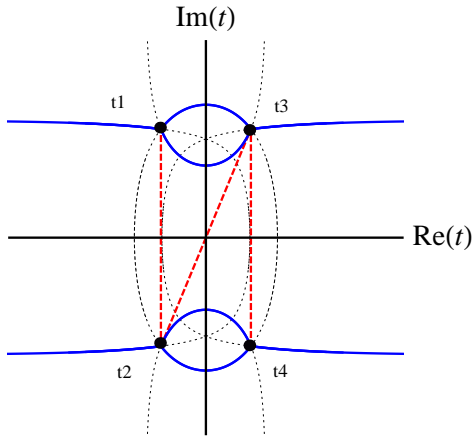}
\caption{The 4 complex turning points,  $t_1, \dots, t_4$, for the field (\ref{evenfield}), showing also the anti-Stokes  lines (solid, blue, lines), Stokes lines (dotted, black, lines) and the integration contours (dashed, red, lines) used in (\ref{scalar}) and (\ref{spinor}).}
\label{stokes}
\end{figure}

\begin{figure}[htb]
\includegraphics[scale=0.55]{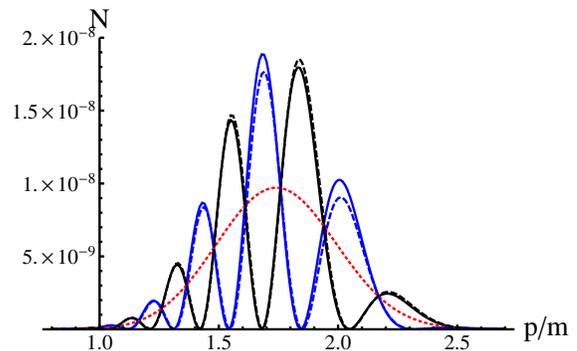}
\caption{Longitudinal momentum spectrum of  $e^+\,e^-$ pairs for the field (\ref{evenfield}), for $E=0.2$, $\omega=0.1$, all in units of $m$. The solid lines are our WKB expressions in (\ref{scalar}) and (\ref{spinor}), the dashed lines are exact numerical results, and the dotted (red) line is the na\"ive ITM expression,  neglecting the interference term. The oscillatory blue lines are scalar QED and the oscillatory black lines are spinor QED. The quantitative agreement of (\ref{scalar}) and (\ref{spinor}) with the numerics is excellent.}
\label{spectrum}
\end{figure}

To illustrate the oscillatory phenomenon most clearly we consider a simple analytic profile field that exhibits the effect found in \cite{Hebenstreit:2009km}. The maximal oscillation effect occurs with carrier-phase $\phi=\pi/2$, so that $E(t)$ is an odd function, and $A(t)$  an even function, so we choose
\begin{eqnarray}
A(t)=\frac{E/\omega}{(1+\omega^2 \, t^2)}
\label{evenfield}
\end{eqnarray}
The scattering potential is plotted in Fig. \ref{potential}. The algebraic form of the vector potential makes it easy to find the turning points, occurring as two complex conjugate pairs: $t_1(p)=-\frac{\sqrt{E/\omega-(p+i m) }}{\omega\sqrt{p+i m}}=t_2(p)^*=-t_3(p)^*=-t_4(p)$, as shown in Fig. 2. Notice that all four t.p.'s are equidistant from the real axis, for all $p$. Fig. \ref{stokes} also shows the Stokes and anti-Stokes lines. Since there are two pairs of  t.p.'s, the WKB analysis leading to (\ref{onepair}) and (\ref{gaussian}) must be generalized to account for the crossing of multiple Stokes and anti-Stokes lines for multiple pairs of t.p.'s in evolving from $t=-\infty$ to $t=+\infty$. This corresponds to the case of over-the-barrier scattering with two bumps in the scattering potential, which has been solved in \cite{froman} using the phase integral approximation (PIA). 
Adapting their result, we find the simple expression
\begin{eqnarray}
N_{\rm scalar}&\approx &e^{-2K_1}+e^{-2 K_2}+ 2\cos(2\alpha)\,e^{-K_1-K_2}
\label{scalar}\\
K_1&=&\left| \int_{t_1}^{t_2} Q\, dt\, \right | \quad,\quad
K_2=\left| \int_{t_3}^{t_4} Q\, dt\, \right | \nonumber\\
\alpha&=& L-\sigma(K_1) -\sigma(K_2) \nonumber \\
L &=&\left| {\mathcal Re}\left( \int_{t_2}^{t_3} Q\, dt\right) \right | \nonumber\\
\sigma(K)&=&\frac{1}{2}\left[\frac{K}{\pi}\left(\ln \left(\frac{K}{\pi}\right)-1\right)+{\rm Arg}\,\Gamma\left(\frac{1}{2}-i\frac{K}{\pi}\right)\right]\nonumber
\end{eqnarray}
In fact, in this case $K_1=K_2$, so we can write $N_{\rm scalar}\approx 4\cos^2(\alpha)e^{-2K_1}$. Note the appearance in (\ref{scalar}) of the interference term, $\cos(2\alpha)$, involving an integral between different pairs of turning points. This term is responsible for the oscillations in the momentum spectrum, as shown in Fig. \ref{spectrum} where we compare (\ref{scalar}) with the exact numerical result, and with a na\"ive application of the ITM result (\ref{onepair}), just taking the first two terms in (\ref{scalar}). The agreement of (\ref{scalar}) with the numerical result is excellent. A generalization to more than two pairs of t.p.'s is discussed in \cite{meyer}.

Having given a quantitative semiclassical explanation of the longitudinal momentum oscillations for scalar QED, we now turn to spinor QED, for which there is a similar scattering formulation \cite{popov,kimpage,Dumlu:2009rr}. The key difference is that the unitarity conditions on the spinor Bogoliubov coefficients have a reversed sign  relative to the scalar case: now $|\alpha|^2+|\beta|^2=1$. This changes the form of the F-matrix of the PIA in \cite{froman}, and is ultimately related to the double-valuedness of the spinor wavefunction. We find the scalar result (\ref{scalar}) is modified to 
\begin{eqnarray}
N_{\rm spinor}&\approx &e^{-2K_1}+e^{-2 K_2}- 2\cos(2\alpha)\,e^{-K_1-K_2}
\label{spinor}
\end{eqnarray}
where the $K_i$ and $\alpha$ are defined as in (\ref{scalar}).  The only change is the sign of the interference term. (When $K_1=K_2$, as for the field in (\ref{evenfield}), we have $N_{\rm spinor}\approx 4\sin^2(\alpha)\,e^{-2K_1}$.) Physically, this term is produced by interference between waves reflected by the double-bump structure, and for fermions there is an additional phase shift on reflection, which ultimately leads to this sign change. In Fig. \ref{spectrum} we plot this spinor result (\ref{spinor}) and we see that it is in excellent agreement with the exact numerical results. The results (\ref{scalar}) and (\ref{spinor})  explain clearly why the oscillations are out of phase between spinor and scalar QED, and why the envelope of the two is the na\"ive ITM result. Of course, if one is interested only in the {\it total} pair production rate, obtained by integrating over $p$, then the difference between spinor and scalar QED is washed out, and agrees with the answer obtained by integrating over the envelope result coming from just the first two terms in (\ref{scalar}) or (\ref{spinor}), since they oscillate about the same envelope.
\begin{figure}[htb]
\includegraphics[scale=0.65]{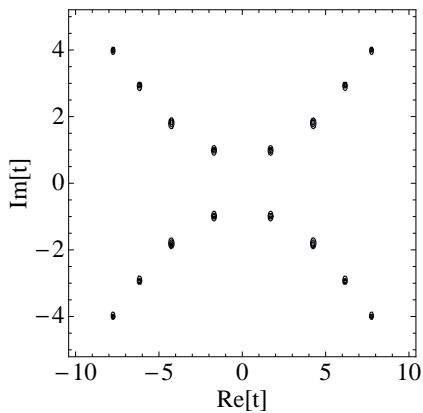}
\caption{Contour plot of $| m^2+(p-A(t))^2|$ for the electric field (\ref{eqn:elfield}), with $\phi=\pi/2$, $p=0$ and $\omega\tau=4$, showing the infinite set of pairs of complex turning points. The two central pairs closest to the real axis  dominate the semiclassical analysis.}
\label{contours}
\end{figure}
To conclude, we sketch in Fig. \ref{contours} the turning points of the carrier-phase field in (\ref{eqn:elfield}), for the case $\phi=\pi/2$. Note that there are infinitely many pairs of complex conjugate t.p.'s. But the two pairs of t.p.'s closest to the real axis dominate, and using formulas (\ref{scalar}) and (\ref{spinor}), we reproduce the oscillatory behavior of the electron-positron longitudinal momentum spectrum found numerically in  \cite{Hebenstreit:2009km}. When $\phi=0$ the sensitivity to the value of $\omega \tau$ can also be understood in terms of the location of the t.p.'s.

Our result is general and simple to use, and has applications beyond this particle-production context, for example to strong-field ionization of atoms and molecules \cite{mourou,popov-ionization}, to particle production problems in cosmology \cite{KeskiVakkuri:1996gn,leonhardt}  and to quasinormal modes of  black holes \cite{Berti:2009kk}. The basic message is that when the time dependence of the external background field has more substructure than just a single bump, the usual textbook ITM result (\ref{onepair}) generalizes in an interesting way that requires fuller consideration of the Stokes line structure of the associated scattering problem, and there are important differences between the momentum spectra of produced spinor or scalar particles. This also has important implications for  the worldline approach to pair production \cite{wli}, which has the potential to describe pair production in electric fields with both temporal and spatial inhomogeneity.


We  acknowledge the DOE  grant DE-FG02-92ER40716, 
and thank C. Bender for bringing \cite{heading} to our attention.


\begin{thebibliography}{999}




\bibitem{eh}
W. Heisenberg and H. Euler,
``Consequences of Dirac's Theory of Positrons'',
Z. Phys. {\bf 98}, 714 (1936);
English translation at arXiv:physics/0605038.

\bibitem{schwinger}
J.~Schwinger, 
``On gauge invariance and vacuum polarization'', 
Phys. Rev. {\bf 82} (1951) 664.

\bibitem{dunne-eli}
G.~V.~Dunne,
``New Strong-Field QED Effects at ELI: Nonperturbative Vacuum Pair Production,'' 
Eur. Phys. J. D {\bf 55}, 327 (2009)
[arXiv:0812.3163 [hep-th]].

\bibitem{tajima}
T.~Tajima, 
``Prospect for extreme field science'',
Eur. Phys. J. D {\bf 55}, 519 (2009).

\bibitem{Schutzhold:2008pz}
  R.~Sch\"utzhold, H.~Gies and G.~Dunne,
  ``Dynamically assisted Schwinger mechanism,''
  Phys.\ Rev.\ Lett.\  {\bf 101}, 130404 (2008)
  [arXiv:0807.0754 [hep-th]];
  G.~V.~Dunne, H.~Gies and R.~Sch\"utzhold,
  ``Catalysis of Schwinger Vacuum Pair Production,''
  Phys.\ Rev.\  D {\bf 80}, 111301 (2009)
  [arXiv:0908.0948 [hep-ph]].
  
  \bibitem{Bell:2008zzb}
  A.~Bell and J.~Kirk,
  ``Possibility of Prolific Pair Production with High-Power Lasers,''
  Phys.\ Rev.\ Lett.\  {\bf 101}, 200403 (2008).


\bibitem{DiPiazza:2009py}
  A.~Di Piazza, E.~Lotstedt, A.~I.~Milstein and C.~H.~Keitel,
 ``Barrier control in tunneling $e^+ e^-$ photoproduction,''
  Phys.\ Rev.\ Lett.\  {\bf 103}, 170403 (2009)
  [arXiv:0906.0726 [hep-ph]].
  
  \bibitem{Monin:2009aj}
  A.~Monin and M.~B.~Voloshin,
``Photon-stimulated production of electron-positron pairs in electric field,''
  Phys.\ Rev.\  D {\bf 81}, 025001 (2010)
  [arXiv:0910.4762 [hep-th]];
and ``Semiclassical Calculation of Photon-Stimulated Schwinger Pair Creation,'' arXiv:1001.3354 [hep-th].

\bibitem{Heinzl:2010vg}
  T.~Heinzl, A.~Ilderton and M.~Marklund,
  ``Finite size effects in stimulated laser pair production,''
 arXiv:1002.4018 [hep-ph].
  
  \bibitem{Bulanov:2010ei}
  S.~S.~Bulanov, V.~D.~Mur, N.~B.~Narozhny, J.~Nees and V.~S.~Popov,
  ``Multiple colliding electromagnetic pulses: a way to lower the threshold of
  $e^+e^-$ pair production from vacuum,''
  arXiv:1003.2623 [hep-ph].

  
\bibitem{ralf-eli}
R. Sch\"utzhold,  C. Maia, 
``Quantum radiation by electrons in lasers and the Unruh effect'',
Eur. Phys. J. D {\bf 55}, 375 (2009).

  \bibitem{leonhardt}
  U.~Leonhardt, T.~Kiss and P.~\"Ohberg, 
  J. Opt. B {\bf 5}, S42 (2003).
  
\bibitem{mourou} 
 G.~Mourou, T.~Tajima and S.~Bulanov,  
``Optics in the relativistic regime,'' 
Rev.\ Mod.\ Phys.\  {\bf 78}, 309 (2006);
M.~Marklund and P.~Shukla,  
``Nonlinear collective effects in photon photon and photon plasma 
interactions,''  
Rev.\ Mod.\ Phys.\  {\bf 78}, 591 (2006) 
[arXiv:hep-ph/0602123];
Y.~Salamin, S.~Hu, K.~Hatsagortsyan and C.~Keitel,  
``Relativistic high-power laser-matter interactions,''  
Phys.\ Rept.\  {\bf 427}, 41 (2006).

\bibitem{astro}
B.~A.~Remington, R.~P.~Drake and D.~D.~Ryutov, 
``Experimental astrophysics with high power lasers and Z pinches'',
 Rev.\ Mod.\ Phys.\ {\bf 78}, 755 (2006).
 
 \bibitem{dima}
  D.~Kharzeev, E.~Levin and K.~Tuchin,
  ``Multi-particle production and thermalization in high-energy QCD,''
  Phys.\ Rev.\  C {\bf 75}, 044903 (2007)
  [arXiv:hep-ph/0602063].



\bibitem{Hebenstreit:2009km}
F.~Hebenstreit, R.~Alkofer, G.~V.~Dunne and H.~Gies,
  ``Momentum signatures for Schwinger pair production in short laser pulses
  with sub-cycle structure,''
  Phys.\ Rev.\ Lett.\  {\bf 102}, 150404 (2009)
   [arXiv:0901.2631 [hep-ph]]; and 
  ``Quantum statistics effect in Schwinger pair production in short laser pulses,''
 arXiv:0910.4457 [hep-ph].
 

 \bibitem{brezin}
E.~Br\'ezin and C.~Itzykson,
``Pair Production In Vacuum By An Alternating Field,''
Phys.\ Rev.\ D {\bf 2}, 1191 (1970).


\bibitem{popov}
V.~S.~Popov,
``Pair Production in a Variable External Field (Quasiclassical approximation)'', 
Sov. Phys. JETP {\bf 34}, 709 (1972); 
``Pair production in a variable and homogeneous electric field as an oscillator problem''.
Sov. Phys. JETP {\bf 35}, 659 (1972);
M.~S.~Marinov and V.~S.~Popov,
  ``Electron-Positron Pair Creation From Vacuum Induced By Variable Electric Field,''
  Fortsch.\ Phys.\  {\bf 25}, 373 (1977).

\bibitem{popov-ionization}
V.~S.~Popov, 
``Tunnel and multiphoton ionization of atoms and ions in a strong laser field (Keldysh theory)''
Phys. Usp. {\bf 47}, 855 (2004);
``Imaginary Time Method in Quantum Mechanics and Field Theory'', 
Phys. Atom. Nucl. {\bf 68}, 686 (2005);
B.~M.~Karnakov, V.~D.~Mur, S.~V.~Popruzhenko and V.~S.~Popov, 
``Strong field ionization by ultrashort laser pulses: application of the Keldysh theory'',
Phys.\ Lett.\ A{\bf 374}, 386 (2009).


 

 \bibitem{kimpage}
 S.~P.~Kim and D.~Page,
  ``Schwinger pair production via instantons in a strong electric field,''
  Phys.\ Rev.\ D {\bf 65}, 105002 (2002)
  [arXiv:hep-th/0005078],
  ``Improved approximations for fermion pair production in inhomogeneous electric fields,''
  Phys.\ Rev.\  D {\bf 75}, 045013 (2007)
  [arXiv:hep-th/0701047].

  \bibitem{Kleinert:2008sj}
  H.~Kleinert, R.~Ruffini and S.~S.~Xue,
  ``Electron-Positron Pair Production in Space- or Time-Dependent Electric
  Fields,''
  Phys.\ Rev.\  D {\bf 78}, 025011 (2008)
  [arXiv:0807.0909 [hep-th]].


  
  
 \bibitem{doubleslit}
 P.~Szriftgiser, D.~Gu\'ery-Odelin, M.~Arndt and J.~Dalibard, 
 ``Atomic Wave Diffraction and Interference using Temporal Slits'',
 Phys.\ Rev.\ Lett.\ {\bf 77}, 4 (1996);
 F.~Lindner et al,
 ``Attosecond Double-Slit Experiment'',
  Phys.\ Rev.\ Lett.\ {\bf 95}, 040401 (2005).
  
  \bibitem{paulus}
  G.~G.~Paulus et al,
  ``Above-threshold ionization by an elliptically polarized field: quantum tunneling interferences and classical dodging'',
  Phys.\ Rev.\ Lett.\ {\bf 80}, 484 (1998).
  
  \bibitem{king}
  B.~King, A. Di Piazza and C.~H.~Keitel,
  ``A matterless double slit'',
 Nature Photon. {\bf 4}, 92 (2010);
 M.~Marklund,
 ``Fundamental optical physics: Probing the quantum vacuum'',
 Nature Photon. {\bf 4}, 72 (2010).
 
  \bibitem{kluger}
  Y.~Kluger, J.~Eisenberg, B.~Svetitsky, F.~Cooper and E.~Mottola,
  ``Pair production in a strong electric field,''
  Phys.\ Rev.\ Lett.\  {\bf 67}, 2427 (1991);
Y.~Kluger, E.~Mottola and  J.~Eisenberg,
  ``The quantum Vlasov equation and its Markov limit,''
Phys.\ Rev.\  D {\bf 58}, 125015 (1998)
  [arXiv:hep-ph/9803372].

\bibitem{Dumlu:2009rr}
  C.~K.~Dumlu,
  ``On the Quantum Kinetic Approach and the Scattering Approach to Vacuum Pair Production,''
  Phys.\ Rev.\  D {\bf 79}, 065027 (2009)
  [arXiv:0901.2972 [hep-th]].

  
  \bibitem{heading}
  J.~Heading, {\it An Introduction to Phase-Integral Methods}, (Methuen, London, 1962).
  
  
    \bibitem{Berry:1972na}
  M.~V.~Berry and K.~E.~Mount,
  ``Semiclassical Approximations In Wave Mechanics,''
  Rept.\ Prog.\ Phys.\  {\bf 35}, 315 (1972).
  
  \bibitem{bender}
  C.~M.~Bender and S.~A.~Orszag, {\it Advanced Mathematical Methods for Scientists and Engineers}, (Springer, 1999).
  
  \bibitem{white}
  R.~White, {\it Asymptotic Analysis of Differential Equations}, (Imperial College Press, 2006).
  
\bibitem{dunnehall}
G.~V.~Dunne, T.~Hall,
  ``On the QED effective action in time dependent electric backgrounds,''
  Phys.\ Rev.\  D {\bf 58}, 105022 (1998)
  [arXiv:hep-th/9807031].
  
 \bibitem{florian-1}
F.~Hebenstreit, R.~Alkofer and H.~Gies,
 ``Pair Production Beyond the Schwinger Formula in Time-Dependent Electric
  Fields,''
  Phys.\ Rev.\  D {\bf 78}, 061701 (2008)
  [arXiv:0807.2785 [hep-ph]].


\bibitem{landau} L.~D.~Landau and L.~M.~Lifshitz, 
{\it Quantum Mechanics (Nonrelativistic Theory)}, (Pergamon, 2003).


\bibitem{pokrovskii}
V.~L.~Pokrovskii and I.~M.~Khalatnikov, 
``On the problem of above-barrier reflection of high-energy particles'', 
Sov. Phys. JETP {\bf 13}, 1207 (1961).

 \bibitem{popov-pulse}
  V.~S.~Popov,
  ``On Schwinger mechanism of e+ e- pair production from vacuum by the  field of optical and X-ray lasers,''
  JETP Lett.\  {\bf 74}, 133 (2001)
  [Pisma Zh.\ Eksp.\ Teor.\ Fiz.\  {\bf 74}, 151 (2001)].


   \bibitem{froman}
  N. Fr\"oman, \"O.~Dammert,
  ``Tunneling and super-barrier transmission through a system of two real potential barriers'',
  Nucl. Phys. {\bf A147}, 627 (1970);
N. Fr\"oman, P. Fr\"oman, {\it Physical Problems Solved by the Phase-Integral Method}, (Cambridge Univ. Press, 2002).


\bibitem{meyer}
R.~E.~Meyer, 
``Quasiclassical scattering above barriers in one dimension'',
J.\ Math.\ Phys.\ {\bf 17}, 1039 (1976).


  \bibitem{KeskiVakkuri:1996gn}
  E.~Keski-Vakkuri and  P.~Kraus,
  ``Tunneling in a Time Dependent Setting,''
  Phys.\ Rev.\  D {\bf 54}, 7407 (1996)
  [arXiv:hep-th/9604151].
  



\bibitem{Berti:2009kk}
N.~Andersson and C.~J.~Howls,
``The asymptotic quasinormal mode spectrum of non-rotating black holes'', 
Class. Q. Grav. {\bf 21}, 1623 (2004);
  E.~Berti, V.~Cardoso and A.~O.~Starinets,
  ``Quasinormal modes of black holes and black branes,''
  Class.\ Quant.\ Grav.\  {\bf 26}, 163001 (2009)
  [arXiv:0905.2975 [gr-qc]].


\bibitem{wli}
  G.~V.~Dunne and C.~Schubert,
  ``Worldline instantons and pair production in inhomogeneous fields,''
  Phys.\ Rev.\ D {\bf 72}, 105004 (2005)
  [arXiv:hep-th/0507174];
  G.~V.~Dunne, Q.-h.~Wang, H.~Gies and C.~Schubert,
   ``Worldline instantons. II: The fluctuation prefactor,''
  Phys.\ Rev.\ D {\bf 73}, 065028 (2006)
  [arXiv:hep-th/0602176].
  

%
  





 

\end{thebibliography}
\end{document}